\documentstyle[aps,prl,twocolumn,psfig]{revtex}
\begin{document}
\preprint{\vbox{\hbox{UCB-PTH-99/52}},
  \vbox{LBNL-44596}}
\draft
\wideabs{
\title{Exponentially Small Supersymmetry Breaking from Extra Dimensions}
\author{Nima Arkani-Hamed, Lawrence Hall, David Smith and Neal Weiner}
\address{
Department of Physics,
University of California,
Berkeley, CA~~94720, USA;\\ 
Theory Group,
Lawrence Berkeley National Laboratory,
Berkeley, CA~~94720, USA}
\date{\today}
\maketitle

\begin{abstract}
        The supersymmetric ``shining'' of free massive chiral superfields in
        extra dimensions from a distant source brane can trigger 
        exponentially small supersymmetry breaking on our brane of order $e^{-2\pi 
        R}$, where $R$ is the radius of the extra dimensions. This supersymmetry 
        breaking can be transmitted to the superpartners in a number 
        of ways, for instance by gravity
        or via the standard model gauge interactions.  The radius $R$
        can easily be stabilized at a size $O(10)$ larger that the fundamental
        scale.  The models are extremely simple, relying only on free,
        classical bulk dynamics to solve the hierarchy problem.

\end{abstract}
}
\setcounter{footnote}{0}
\setcounter{page}{1}
\setcounter{section}{0}
\setcounter{subsection}{0}
\setcounter{subsubsection}{0}


%
%
%



\noindent{\bf 1} The four forces of nature are each characterized by a mass scale: 
$\sqrt{1/G_N} = M_{P} \approx 10^{19}$ GeV for gravity, $\Lambda_W 
\approx 10^3$ GeV for the weak interaction, $\Lambda_{QCD} \approx 0.1$ 
GeV for the strong interaction and $m_\gamma = 0$ for the electromagnetic 
interaction. What is the origin of these diverse scales? Over the last 25 
years a single dominant viewpoint has developed: the largest scale, that 
of gravity, is fundamental, and the other scales are generated by a 
quantum effect in gauge theories known as dimensional transmutation. If 
the coupling strengths of the other forces have values $\alpha_{P} 
\approx 1/30$ at the fundamental scale, then a logarithmic evolution of 
these coupling strengths with energy leads, in non-Abelian theories, to the 
generation of a new mass scale 
\begin{equation}
\Lambda \approx M_{P} \; e^{-1 / \alpha_{P}}
\label{eq:DT}
\end{equation}
where the interaction becomes non-perturbative. On the 
other hand, Abelian theories, like QED, remain perturbative to 
arbitrarily low scales. For strong and electromagnetic interactions this 
viewpoint is immediately successful; but for the weak interaction the 
success is less clear, since the weak interactions are highly perturbative 
at the scale $\Lambda_W$. If $\Lambda_W$ is generated by a dimensional 
transmutation, it must happen indirectly by some new force getting strong 
and triggering the breakdown of electroweak symmetry. There have been 
different 
ideas about how this might occur: the simplest idea is technicolor, a 
scaled up version of the strong force\cite{SW}; another possibility has the new 
strong force first triggering supersymmetry breaking which in turn triggers 
electroweak symmetry breaking\cite{witten}. For our purposes the crucial thing about 
these very different schemes is that they have a common mechanism 
underlying the origin of $\Lambda_W$: a dimensional 
transmutation, caused by the logarithmic energy evolution of a gauge 
coupling constant, generates the 
exponential hierarchy of (\ref{eq:DT}).

In this letter,
we propose an alternative mechanism for generating $\Lambda_W$ 
exponentially smaller than the fundamental scale. Our scheme requires 
two essential ingredients beyond the standard 
model: supersymmetry, and 
compact extra dimensions of space. The known gauge 
interactions reside on a 3-brane, and physics of the surrounding bulk 
plays a crucial role in generating an exponentially small scale of 
supersymmetry breaking.

Our mechanism is based on the idea of ``shining'' \cite{AD}. 
A bulk scalar field, $\phi$, of mass $m$, 
is coupled to a classical source, $J$, on a brane at location
$y=0$ in the bulk, thereby acquiring an exponential profile 
$\phi \propto J e^{-m|y|}$ in all regions of the bulk distant
from the source, $m |y| \gg 1$. If our brane is distant from the
source, then this small exponential, arising from the propagation 
of the heavy scalar across the bulk, can provide an origin for very small
dimensionless numbers on our brane, in particular for supersymmetry 
and electroweak symmetry breaking
\begin{equation}
\Lambda_W \propto M_* \; e^{-mR}
\label{eq:susyshine}
\end{equation}
where $R$ is the distance scale of our brane from the source brane, and 
$M_*$ is the fundamental scale of the theory. 
The possibility of such a supersymmetry-breaking mechanism has been 
noted before qualitatively \cite{AD}. If some of the 
extra dimensions are very large, $M_*$ can be significantly below $M_P$, 
and could even be of order $\Lambda_W$, providing an alternative 
viewpoint on the mass scales of the four fources of nature \cite{ADD}. 
We are concerned with the case of $M_* \gg \Lambda_W$, although $M_*$ need 
not be as large as $M_P$. 
In this letter 
we give an explicit construction of shining which preserves 
4-dimensional supersymmetry, but triggers an exponentially small 
amount of supersymmetry breaking due to 
the presence of our brane.  
A possible worry is that $R$ might run to infinity, thus 
minimizing the vacuum energy and restoring supersymmetry. 
We exhibit simple mechanisms, based on the same supersymmetric 
shining, which stabilize the extra dimensions with finite radius.


\noindent{\bf 2} 
We begin by constructing a 5d theory, with a source brane shining an
exponential profile for a bulk scalar,
such that the equivalent 4d theory
is exactly supersymmetric. The 5d theory possesses N=1
supersymmetry in a representation containing two scalar fields, $\phi$ and 
$\phi^{c}$, together with a four-component spinor $\Psi = (\psi, 
\psi^{c})$. The equivalent 4d theory has
two families of chiral superfields $\Phi(y) = \phi(y) + \theta \psi(y) +
\theta^2 F(y)$ and 
$\Phi^{c}(y) = \phi^c(y) + \theta \psi^c(y) + \theta^2 F^c(y)$. 
In the 4d theory, $y$ can be viewed as a parameter labelling the families of
chiral superfields.


Using this 4d chiral superfield notation, we write the bulk action as
\begin{eqnarray}
S_B=\int d^{4}x\>dy &\bigl(& \int d^4 \theta (\Phi^\dagger \Phi 
+ \Phi^{c\dagger}
\Phi^c) \nonumber \\ 
&+& \int d^2 \theta \Phi^c (m+ \partial_y) \Phi \bigr)
\label{eq:bulklag}
\end{eqnarray}
Viewed as a 4d theory, we have manifest supersymmetry,
with the $y$ integral summing over the family
of chiral superfields.
The form of the superpotential appears somewhat
unusual; however, on eliminating the auxiliary fields, the action in
terms of component fields describes a free Dirac fermion and two 
complex scalar fields in 5d.
The 5d Lorentz invariance is not
manifest in (\ref{eq:bulklag}), but this form is useful to us, since
it makes the 4d supersymmetry manifest. 

Next we locate a 3-brane at $y=0$, and require that it provides a
source, $J$, for a chiral superfield in a way which
preserves 4d supersymmetry: 
\begin{equation}
W_S = \int dy \delta(y) \>J \Phi^c  ,
\label{eq:linsource}
\end{equation}
where we choose units so that the fundamental scale of the theory $M_*=1$. 
The conditions that this source shines scalar fields into the bulk
such that supersymmetry is not spontaneously broken are
\begin{eqnarray}
F(y) &=& (m-\partial_y)\phi^c = 0 \\
F^c(y) &=& J\delta(y)+(m+\partial_y) \phi = 0
\label{eq:fflatness}
\end{eqnarray}
The first of these does not have any non-trivial solutions that do not
blow up at infinity, or which are well-defined on a circle. The second, however,
has the solution
\begin{equation}
\phi(y) = - \theta(y) J e^{- m y},
\label{eq:phiprofile}
\end{equation}  
in infinite flat space and
\begin{equation}
\phi(y) = {-J e^{-my} \over 1-e^{-2 \pi m
    R}} \hskip 0.15in y \in [0,2 \pi R) ,
\label{eq:phiprofile2}
\end{equation}  
on a circle.
%
Thus we see that $\phi$ has taken on a non-zero profile in the
bulk, but in a way that the energy of the system remains zero and
one supersymmetry remains unbroken.
Interestingly, this is not the profile that occurs with non-supersymmetric
shining, but is asymmetric, shining in only one direction. 
One may have thought that the gradient energy for any profile of a
bulk scalar field would neccessarily break supersymmetry, but our example shows 
this is not the case.  
The $|F^c|^2$ contribution to the vacuum energy includes the $|\partial_y 
\phi|^2 + |m \phi|^2$ terms as expected, but these are cancelled by
$\phi^* \partial_y \phi$ terms, and at $y=0$ by terms which arise
because $J$ is coupled to the combination $(m+ \partial_y) \phi(0)$.
Note that
if we had written a linear term for $\Phi$ instead of
$\Phi^c$, we would have shined a profile for $\phi^c$ in the opposite
direction. Likewise, if we had chosen a negative value for $m$, we
would shine $\phi$ in the opposite direction, since the 5d theory is
invariant under $m \rightarrow -m, \;\; y \rightarrow -y$.

\noindent{\bf 3}
Having learned how to shine a chiral superfield from a source brane 
across the bulk, we now investigate whether a probe brane, located 
far from the source at $y=\overline y$, can sample the small value 
of $\phi(\overline y)$ to
break supersymmetry by an exponentially small amount on the probe brane.
In addition to  superfields which contain the standard model fields,
the probe brane contains a standard model singlet chiral superfield
$X$, and has a superpotential
\begin{eqnarray}
W_P = \int dy \>  \delta(y-{\overline y}) ( W_{MSSM} + \Phi X)
\label{eq:nosusyw}
\end{eqnarray}
where $W_{MSSM}$ is the superpotential of the minimal supersymmetric
standard model.
This superpotential has F-flatness conditions
\begin{eqnarray}
F^c(y) &=& J \delta(y) + (m+\partial_y)\phi =0 \\
F(y) &=& \delta(y-\overline y) x + (m-\partial_y) \phi^c \\
F_{X} &=&  \phi(\overline{y}). 
\label{eq:nonsusyfs}
\end{eqnarray}
The first equation can only be satisfied by 
having a shined value for $\phi(\overline y) \ne 0$.
Clearly, the first and third equations cannot be simultaneously
satisfied: we have an O'Raifeartaigh theory,
and supersymmetry is spontaneously broken.
As always in an O'Raifeartaigh theory, at tree level there is a flat 
direction: the value for $x$ is undetermined, and if it is non-zero it
acts as a source shining $\phi^c$. 
It is simple to understand what is going on. In the presence of the
source brane, the field $\phi$ is shined from the source brane,
generating an exponentially small linear term for $X$ on the probe
brane. After we have integrated out the heavy fields $\phi$
and $\phi^c$ we are simply left with the superpotential on the
probe brane
\begin{equation}
W_P \sim  J e^{- m \overline y} X,
\label{eq:susybreaking}
\end{equation}
which generates a nonzero $F_X \sim J e^{-m \overline y}$.

This is not a precise equality, 
as the probe brane resists a non-zero $\phi(\overline y)$, and provides a
back reaction on the bulk. It is simple to show that this effect is 
qualitatively insignificant.

\begin{figure}
  \centerline{
    \psfig{file=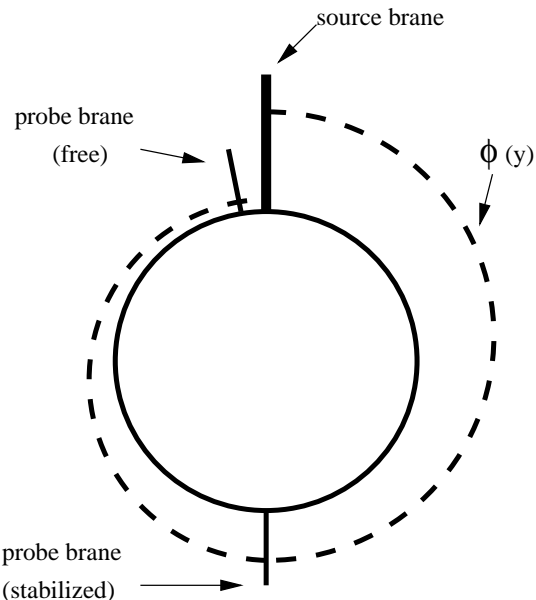,width=0.4\textwidth,angle=0} }
        \caption[sniff]{The schematic profile of $\phi$ in the
          extra dimension. Whether our brane is stabilized at some
          position or free to move
          under the given forces, we can achieve an exponentially
          small value for $\phi$ and hence exponentially suppressed
          supersymmetry breaking.}
        \label{fig:susyplot}
\end{figure}

If the fifth dimension is a circle, then we can imagine that the 
probe brane is stabilized at some location on the circle, or that it will
drift such that it is immediately next to the source brane where the
resulting supersymmetry
breaking is smallest, as in figure \ref{fig:susyplot}. In either case, 
we generate an exponentially small supersymmetry breaking scale
$F_X$. 

Notice that this is {\it not} in the same spirit as recent works that 
use bulk dynamics to transmit distantly broken 
supersymmetry\cite{RSetal}. Rather, in our case, in the absence of 
either source or probe brane, supersymmetry remains unbroken. It is 
the simultaneous presence of {\it both} branes that leads to the 
exponentially small supersymmetry breaking.
A simple option for mediating the supersymmetry breaking from $F_X$ to
the standard model superpartners is to add non-renormalizable
operators to the probe brane
\begin{eqnarray}
\Delta S_P = \int d^4x dy \delta(y & - & \overline y) \bigl( \int d^4
\theta( {1 \over M_*^2} X^\dagger X Q^\dagger Q + ... ) \nonumber \\
& + & \int d^2 \theta ( { 1 \over M_*} X W^\alpha W_\alpha + ...) 
\bigr)
\label{eq:higherdim}
\end{eqnarray}
where $Q$ is a quark superfield and $W^\alpha$ a standard model gauge
field strength superfield. We have inserted $M_*$ explicitly, so that
the soft masses of the standard model superpartners and 
$x$ are $\widetilde{m} \sim
F_X/M_* \sim(J/M_*)e^{-m \overline{y}}$. 
Until now we have not specified the values for $J$ and $m$; the most
natural values are $J \approx M_*^2$ and $m \approx M_*$. 

Our entire theory is remarkably simple, and is specified by the bulk
action $S_B$ of (\ref{eq:bulklag}), the source brane superpotential $W_S$ of
(\ref{eq:linsource}), and the interactions of (\ref{eq:nosusyw}) and 
(\ref{eq:higherdim}) on our brane.

\noindent{\bf 4} 
Mechanisms for dynamical supersymmetry
breaking by dimensional transmutation\cite{ADS} typically suffer from the 
``dilaton runaway problem'' when embedded in string theory\cite{DS}:  
since the coupling constant $\alpha_{P}$ is a dynamical field, the vacuum 
energy is minimized as $\alpha_{P} \rightarrow 0$, where the theory 
becomes free.  
In our case, it appears there is an analogous problem.
Taking the supersymmetry-breaking 
brane to be free to drift, the vacuum energy of the theory is
\begin{equation}
E \sim  J^2 e^{-4 \pi R m},
\label{eq:VACEN}
\end{equation}
so it is energetically favorable for the radius to grow to
infinity. 
However, in contrast with dynamical supersymmetry breaking 
scenarios,  where one must simply assume that the dilaton
vev is somehow prevented from running to infinity, 
stabilizing $R$ turns out to be quite simple.

Consider adding to the model of the previous section a second bulk
multiplet $(\Phi' , \Phi'^{c})$, of mass $m'$, with interactions
\begin{equation}
W' = \int \! dy \, \left[  \delta(y) J' \Phi'^c
    + \delta(y-\overline{y}) X' \left( \Phi' + A \right)\right]
\end{equation}
where $A$ and $J'$ are constants and $X'$ is a chiral 
superfield.  The terms in this superpotential are nearly identical to 
those  
of (\ref{eq:nosusyw}) and (\ref{eq:linsource}), except for the
presence of the constant $A$ on the probe brane.
We assume that both $A$ and $J'$ are real.
In complete analogy with the shining of $\phi$, the scalar $\phi'$
acquires a profile
\begin{equation}
\phi'(y) = -J' \theta (y) e^{- m' y}.
\end{equation}
Writing ${\overline{y}} = \theta R$, the F-flatness condition for 
$X'$ becomes
\begin{equation}
m' R \theta = \log{J' \over A (1-e^{-2 \pi R
    m'}) },
\label{eq:Fflat}
\end{equation}
which defines a real function $R(\theta)$ provided that $J'/A>0$.
We assume $m'$ is less
than $m$ (by a factor of roughly 30, for very large $M_*$),
so that, for a given value of $\theta$, the
radius is essentially determined
by the condition $F_{X'}=0$, with a small correction ${\Delta R \over R}
\sim {m \over m'} e^{-m/m'}$ coming from the $|F_X|^2$
contribution to the potential.  However, we have already seen that the 
vacuum energy is minimized when the probe brane drifts
completely around the circle.  The value 
of $R$ is thus immediately fixed by equation
(\ref{eq:Fflat}), with $\theta = 2\pi$.  Its precise value depends on $A$ and 
  $J'$,  but if we take their ratio to be of order unity, 
then we find $2 \pi R m' 
  \sim 1$.  The supersymmetry
breaking F-term is  then $F_X \sim J e^{-2 \pi m R }\sim
J e^{-m/ m'}$, so that the higher dimension interactions of 
(\ref{eq:higherdim}) give superpartner masses
\begin{equation}
\widetilde{m} \sim e^{-m/m'} M_*.
\end{equation}
In this model the mass of the radion, the field associated with
fluctuations of the size of the circle, is $m_{radion} \sim F_X/M_{P}
\sim$ 1 TeV $(M_*/M_{P})$.

Alternatively one can stabilize R in an entirely supersymmetric
fashion.  Here we describe just one of a number of ways in which this
can be done.  Imagine supplementing the ``clockwise''
shining of $\phi'$ due to $W'$ with ``counterclockwise'' shining of
a different scalar $\widetilde{\phi}^c$ of comparable mass, 
$\widetilde{m}$, through the added
superpotential terms
\begin{equation}
\widetilde{W} = \int \! dy \,\left[\delta(y) \widetilde{J} 
\widetilde{\Phi}  
+   \delta(y-\overline{y}) \widetilde{X} 
\left( \widetilde{\Phi}^{c}  + B \right) \right].
\end{equation}   
Note that because  $\widetilde{\Phi}$ (rather than
${\widetilde{\Phi}}^c$) couples
to the source, the shining is in the opposite direction as that of
$\phi'$.  The F-flatness condition for $\widetilde{X}$,
\begin{equation}
\widetilde{m} R (2\pi  - \theta) = 
\log{B \over \widetilde{J} (1-e^{-2 \pi R
    \widetilde{m}})},
\label{eq:Fflattilde}
\end{equation}
and the F-flatness condition for $X'$ independently determine $R$ as a
function of $\theta$, and for broad ranges of parameters the combined
constraints are satisfied by unique values of $\theta$ and $R$.
This supersymmetric stabilization of the radius yields $m_{radion}
\sim M_*^2/M_{P}$, far above the TeV scale.

\noindent{\bf 5}
We have presented a complete model in which exponentially small
supersymmetry breaking is generated as a bulk effect and
communicated to the standard model via higher-dimension operators. It
is straightforward to modify the model so that the supersymmetry breaking is
mediated instead by gauge interactions\cite{ACW}.

Consider the O'Raifeartaigh superpotential
\begin{equation}
W=X(Y^2-\mu^2)+mZY.
\end{equation}
At tree level $x$ is a flat direction, but provided $\mu^2<m^2/2$, radiative effects stabilize 
$x$ at the origin and give $m_x^2 \sim \mu^2/16 \pi^2$.  
Supersymmetry is broken
by $F_X = -\mu^2$.  Models using an O'Raifeartaigh superpotential to
achieve low-energy supersymmetry breaking have been constructed in the 
past, but have required a small value for $\mu^2$ to be input by
hand.  Instead, we use supersymmetric shining as an
origin for the parameters $\mu^2$ and $m$ by coupling the brane
superfields $X$, $Y$, and $Z$ to the shone $\Phi$ according to
\begin{equation}
W_{hidden}=\lambda_1 X(Y^2-\Phi({\overline y})^2)+\lambda_2
\Phi({\overline y}) ZY,
\label{eq:hidden}
\end{equation}   
where $\lambda_1$ and $\lambda_2$ are both of order unity and
$\lambda_1<\lambda_2^2$/2.
Next we introduce couplings to messenger fields $Q$ and
$\overline{Q}$ transforming under the standard model gauge
group\cite{symmetries},
\begin{equation}
W_{messenger}=\alpha_1 X Q {\overline Q} + \alpha_2 \Phi({\overline y}) Q
{\overline Q}.
\label{eq:messenger}
\end{equation}
By taking $\alpha_2^2>\alpha_1 \lambda_1$ we ensure that the messenger
scalars do not acquire vevs.  These superpotentials give 
$Q$ and $\overline{Q}$ supersymmetric masses and supersymmetry-breaking mass splittings 
of comparable order, $M \sim \sqrt{F} \sim \phi(\overline{y})$.  The
messengers then feed the supersymmetry breaking into the standard model in the usual
way, yielding soft supersymmetry-breaking parameters of order 
$\widetilde{m} \sim {1  \over 16 \pi^2}\phi(\overline{y})$. 
Fixing the radius $R$ by either of the mechanisms already described then leads to
$\widetilde{m} \sim {M_* \over 16 \pi^2}
e^{-m/m'}$.  Note that this is truly a model of low-energy supersymmetry breaking,
with $\sqrt{F}\sim 16\pi^2 \widetilde{m} \sim$ 100 TeV, allowing for decays 
of the NLSP within a detector length.  Moreover, this small value for
$\sqrt{F}$ is favored by cosmology in that it suppresses the gravitino
energy density\cite{dGMM}. 

While there is typically a severe $\mu$ problem in gauge-mediated
theories \cite{DGP}, it is easily solved with our mechanism by
shining $\mu$ in the superpotential with a term
\begin{equation}
W \supset \lambda \phi(\overline y) H_1 H_2.
\label{eq:mu}
\end{equation}
With $\lambda \sim 1/30$, problems of naturalness are much 
less severe than in theories where supersymmetry is broken
dynamically. If $B \mu = 0 $ at tree
level, radiative effects can generate a small
$B \mu$ and large $\tan \beta$ \cite{RatS}. Likewise, in gravity
mediated theories, a shined term $\int d^2 \theta 
\Phi (\overline y) 
H_1 H_2$ can also generate an appropriate value for $\mu$, while $\int
d^4 X^\dagger X H_1 H_2$ generates $B \mu$. Although $\phi$ is
related to supersymmetry breaking, this is distinct from the
Giudice-Masiero mechanism. Absent the superfield $X$, supersymmetry
is preserved, but the value of $\mu$ is unchanged.

Depending on whether supersymmetric or supersymmetry breaking 
stabilization of the radius is employed,
the radion mass is either $m_{radion} \sim M_*^2/M_{P}$ 
or $m_{radion} \sim \sqrt{F}/M_{P} \sim$ 1 eV $(M_*/M_{P})$. 
Even the latter case is safe, 
since the limit on the radion mass is on the mm$^{-1}$ scale, at the limits 
of experimental probes of gravity at short distances.

\noindent{\bf 6}
Dimensional transmutation, (\ref{eq:DT}), and shining, (\ref{eq:susyshine}),
are alternative mechanisms for taking a dimensionless input of order 30 and 
generating an exponentially small mass hierarchy. 
These mass hierarchies can explain the scales of symmetry breaking,
for instance of a global flavor symmetry, or of supersymmetry, as we
have discussed. While dimensional 
transmutation is a quantum effect requiring an initial coupling which 
is highly perturbative, $1/\alpha_P \approx 30$,
shining is classical and requires a bulk distance scale of 
size $R \approx 30 M_{*}^{-1}$.  Such a radius can in turn be stabilized in a 
simple way.  
We presented two standard ways of communicating this 
exponentially small supersymmetry breaking, through higher-dimensional 
operators or via standard model gauge interactions. It is clearly
possible to employ other mechanisms, such as those discussed in \cite{RSetal}. 
Our theories 
are remarkably simple, using only free classical dynamics in one 
extra dimension. Extensions to more dimensions should be 
straightforward.  While we have concentrated on constructing 
effective theories with exponentially small global supersymmetry breaking, it 
will be interesting to embed these models in a consistent local 
supergravity.  It will also be interesting to explore whether any of 
these mechanisms can be realized in the D-brane construction of 
non-BPS states in string theory.
\begin{acknowledgements}
This work was supported in part by the Director, Office of Science,
Office of High Energy and Nuclear Physics, Division of High Energy
Physics of the U.S. Department of Energy under Contract
DE-AC03-76SF00098 and in part by the National Science Foundation under
grant PHY-95-14797. 
\end{acknowledgements}

\end{document}